\documentstyle{mn}
\input epsf.sty
\newif\ifAMStwofonts
\AMStwofontstrue
 
\ifoldfss
  \ifCUPmtlplainloaded \else
    \NewTextAlphabet{textbfit} {cmbxti10} {}
    \NewTextAlphabet{textbfss} {cmssbx10} {}
    \NewMathAlphabet{mathbfit} {cmbxti10} {} 
    \NewMathAlphabet{mathbfss} {cmssbx10} {} 
  \fi
  \ifAMStwofonts
    \ifCUPmtlplainloaded \else
      \NewSymbolFont{upmath} {eurm10}
      \NewSymbolFont{AMSa} {msam10}
      \NewMathSymbol{\upi}     {0}{upmath}{19}
      \NewMathSymbol{\umu}     {0}{upmath}{16}
      \NewMathSymbol{\upartial}{0}{upmath}{40}
      \NewMathSymbol{\leqslant}{3}{AMSa}{36}
      \NewMathSymbol{\geqslant}{3}{AMSa}{3E}

    \fi
  \fi
\fi 
 
\ifnfssone
  \newmathalphabet{\mathit}
  \addtoversion{normal}{\mahit}{cmr}{m}{it}
  \addtoversion{bold}{\mathit}{cmr}{bx}{it}
  \newmathalphabet{\mathbfit} 
  \addtoversion{normal}{\mathbfit}{cmr}{bx}{it} 
  \addtoversion{bold}{\mathbfit}{cmr}{bx}{it}
  \newmathalphabet{\mathbfss} 
  \addtoversion{normal}{\mathbfss}{cmss}{bx}{n}
  \addtoversion{bold}{\mathbfss}{cmss}{bx}{n}
  \ifAMStwofonts
    \ifCUPmtlplainloaded \else
      \UseAMStwoboldmath
      \makeatletter
      \new@mathgroup\upmath@group
      \define@mathgroup\mv@normal\upmath@group{eur}{m}{n}
      \define@mathgroup\mv@bold\upmath@group{eur}{b}{n}
      \edef\UPM{\hexnumber\upmath@group}
      \new@mathgroup\amsa@group
      \define@mathgroup\mv@normal\amsa@group{msa}{m}{n}
      \define@mathgroup\mv@bold\amsa@group{msa}{m}{n}
      \edef\AMSa{\hexnumber\amsa@group}  
      \makeatother
      \mathchardef\upi="0\UPM19
      \mathchardef\umu="0\UPM16
      \mathchardef\upartial="0\UPM40
      \mathchardef\leqslant="3\AMSa36
      \mathchardef\geqslant="3\AMSa3E
    \fi
  \fi
\fi 
   
\ifnfsstwo
  \DeclareMathAlphabet{\mathbfit}{OT1}{cmr}{bx}{it}
  \SetMathAlphabet\mathbfit{bold}{OT1}{cmr}{bx}{it}
  \DeclareMathAlphabet{\mathbfss}{OT1}{cmss}{bx}{n}
  \SetMathAlphabet\mathbfss{bold}{OT1}{cmss}{bx}{n}
  \ifAMStwofonts
    \ifCUPmtlplainloaded \else
      \DeclareSymbolFont{UPM}{U}{eur}{m}{n}
      \SetSymbolFont{UPM}{bold}{U}{eur}{b}{n}
      \DeclareSymbolFont{AMSa}{U}{msa}{m}{n}
      \DeclareMathSymbol{\upi}{0}{UPM}{"19}
      \DeclareMathSymbol{\umu}{0}{UPM}{"16}
      \DeclareMathSymbol{\upartial}{0}{UPM}{"40}
      \DeclareMathSymbol{\leqslant}{3}{AMSa}{"36}
      \DeclareMathSymbol{\geqslant}{3}{AMSa}{"3E}
    \fi
  \fi  
\fi 
      
\ifCUPmtlplainloaded \else
  \ifAMStwofonts \else 
    \def\upi{\pi}
    \def\umu{\mu}
    \def\upartial{\partial}
  \fi
\fi

\title{Black hole masses from power density spectra: determinations and consequences}
\author[B. Czerny, M. Niko\l ajuk, M. Piasecki, J. Kuraszkiewicz]
       {B. Czerny$^1$, M. Niko\l ajuk$^1$, M. Piasecki$^1$, J. Kuraszkiewicz$^2$\\
        $^1$N. Copernicus Astronomical Center, Bartycka 18, 00-716 Warsaw, 
        Poland\\
        $^2$Harvard-Smithsonian Center for Astrophysics, Cambridge, MA 02138, USA}

\begin{document}

\maketitle

\begin{abstract}

We analyze the scaling of the X-ray power density spectra with the
mass of the black hole on the example of Cyg X-1 and Seyfert 1 galaxy
NGC 5548.We show that the high frequency tail of the power density
spectrum can be successfully used for determination of the black hole
mass. We determine the masses of the black holes in 6 Broad Line
Seyfert 1 galaxies, 5 Narrow Line Seyfert 1 galaxies and two QSOs
using available power density spectra. The proposed scaling is clearly
appropriate for other Seyfert galaxies and QSOs. In all but 1 normal
Seyferts the resulting luminosity to the Eddington luminosity ratio is
smaller than 0.15, with a source MCG -6-15-30 being an exception. The
applicability of the same scaling to Narrow Line Seyfert 1 is less
clear and there may be a systematic shift between the power spectra of
NLS1 and S1 galaxies of the same mass, leading to underestimation of
the black hole mass. However, both the method based on variability and
the method based on spectral fitting show that those galaxies have
relatively low masses and high luminosity to the Eddington luminosity
ratio, supporting the view of those objects as analogs of galactic
sources in their high/soft or very high state based on the overall
spectral shape. Bulge masses of their host galaxies are similar to
normal Seyfert galaxies so they do not follow the black hole mass -
bulge mass relation for Seyfert galaxies, being evolutionary less
advanced, as suggested by Mathur (2000).  The bulge mass--black hole
mass relation in our sample is consistent with being linear, with
black hole to bulge ratio $\sim 0.03$ \% , similar to Wandel (1999)
and Laor (1998, 2001) for low mass objects but significantly shifted
from the relation of Magorrian et al. (1998) and McLure \& Dunlop
(2000).

\end{abstract}


\section{Introduction}

Determination of the masses of black holes in active galactic nuclei (AGN)
is a key element in studies of the nature of accretion process and 
evolution of an active nucleus.

Several methods were used so far for this purpose: estimation of the
virial mass from the emission line data through reverberation
techniques (see e.g Wandel 1999), estimation of the virial mass on the
basis of the source luminosity and estimates of the ionization
parameter in Broad Line Region (e.g. Wandel, Peterson \& Malkan 1999),
and model fits (usually of accretion discs) to the broad band spectra
(e.g. Edelson \& Malkan 1986, Sun \& Malkan 1989, Merloni, Fabian \&
Ross 2000).

Suggestion that variability properties of the sources are affected by
their mass and/or luminosity was made by Barr \& Mushotzky (1986) and
Wandel \& Mushotzky (1986), and subsequently explored by many authors
(e.g. McHardy 1989, Green, McHardy \& Lehto 1993, Turner et al. 1999,
Leighly 1999a).

A convenient and powerful method of determination of black hole mass
from variability properties was recently used by Hayashida et
al. (1998) and Hayashida (2000). The method uses the normalization of
the X-ray power density spectra (hereafter PDS) in comparison with Cyg
X-1 in the hard state. The method, however, lead to surprisingly small
values of the black hole masses and, consequently, to luminosity to
the Eddington luminosity ratios frequently larger than one.

In the present work we reconsider the method of Hayashida et al.
(1998). We determine better scaling coefficient on the basis of the
new available power spectra of Cyg X-1, the new power spectrum of NGC
5548, and the independent black hole mass measurement for this AGN.

We determined the black hole masses for a number of AGN with the PDS
available in the literature.  We included both the classical Seyfert 1
galaxies (i.e. Broad Line Seyfert 1 galaxies; hereafter BLS1) and
Narrow Line Seyfert 1 galaxies (NLS1), characterized by much narrower
broad lines ($V_{FWHM} < 2000 $ km s$^{-1}$, Osterbrock \& Pogge
1985).

We discuss the consequences of those black hole mass determination for
the accretion rate in BLS1 and NLS1 galaxies and for their
evolutionary status.

\section{Method}

\subsection{Determination of black hole mass in NGC 5548}
\label{sect:mass5548}

Seyfert 1 galaxy NGC 5548 (z = 0.0174) is one of the most frequently
studied AGN. The mass of the central black hole in this object is
therefore determined relatively accurately. Peterson \& Wandel (1999)
give the value $6.8^{+1.5}_{-1.0} \times 10^7 M_{\odot}$ from
reverberation studies of the various lines emitted by the Broad Line
Region.  This technique may potentially suffer from systematic errors
up to a factor of 3 (Krolik 2000) but recent comparison of statistical
properties of reverberation and dynamical mass measurement indicated
smaller errors (Gebhardt et al. 2000).  Virial mass of the black hole
in NGC 5548, determined on the basis of ionization method for a number
of emission lines is equal to $5.9 \pm 2.5 \times 10^7 M_{\odot}$
(Peterson \& Wandel 2000).

Those measurements confirmed earlier results based on the study of the
BLR which indicated the mass about $10^8 M_{\odot}$ (Wanders et
al. 1995), or somewhat smaller (Done \& Krolik 1996). Rokaki et
al. (1993) argued for the value between $ 5 \times 10^7 M_{\odot}$ and
$ 6 \times 10^7 M_{\odot}$ and Kaspi et al.  (2000) obtained values
$9.4 - 12.3 \times 10^7 M_{\odot}$.

Fits of a simple disc/corona model to the optical/UV continuum data
from the AGN Watch (Clavel et al. 1991, Peterson et al. 1992) favored
the value $10^ 8 M_{\odot}$ although the value $6 \times 10^7
M_{\odot}$ was also acceptable (Loska \& Czerny 1997). Fits of the
more advanced disc/corona model of Witt, Czerny\& \. Zycki (1997) 
which was based also on the X-ray data indicated rather larger value 
$1.4 \times 10^8 M_{\odot}$ (Kuraszkiewicz,  Loska \& Czerny 1997).   

The value of the black hole mass $6.8^{+1.5}_{-1.0} \times 10^7
M_{\odot}$ seems to be determined most accurately, since it resulted
from the analysis of several emission lines coming from the Broad Line
Region. It is also consistent with other determinations, taking into
account quite large and unspecified errors in measurements based on
other methods. 

\subsection{Power density spectrum of NGC 5548 and the comparison 
with galactic sources}
\label{sect:5548vsgbh}

\begin{figure}
\epsfxsize = 90 mm 
\epsfbox{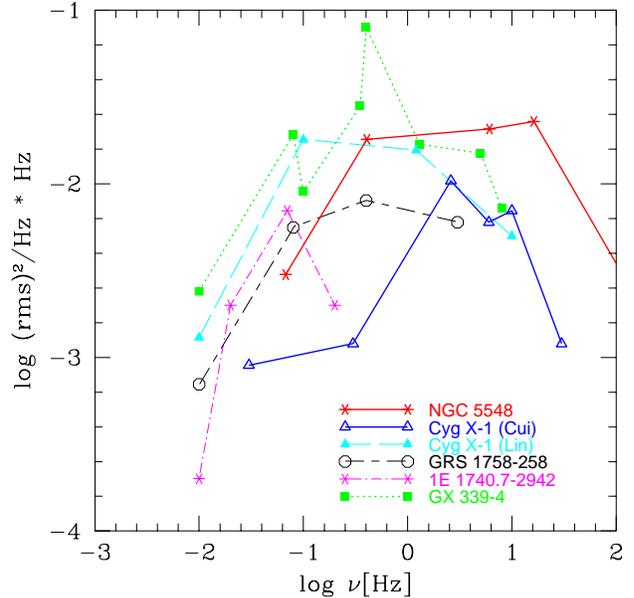}
\caption{Power density spectrum times frequency for NGC 5548 from 
Chiang et al. (2000)
(stars connected with continuous line) and for galactic black holes:
Cyg X-1 in transition state from Cui et al. (1997) (open triangles
connected with continuous line), Cyg X-1 in hard state from Lin et
al. (2000) (solid triangles connected with long dash line), GRS
1758-258 from Lin et al.  (open octagons connected with short
dash-long-dash line), 1E 1740.7-2942 from Lin et al. (stars connected
with dot-short-dash line), GX 339-4 from Lin et al.  (solid squares
connected with dot line).  Power density spectrum of NGC 5548 was
shifted by 6.83 in $log(\nu)$ which corresponds to rescaling this
source to a black hole mass of $10 M_{\odot}$.
\label{fig1}}
\end{figure}

\begin{figure}
\epsfxsize = 90 mm 
\epsfbox{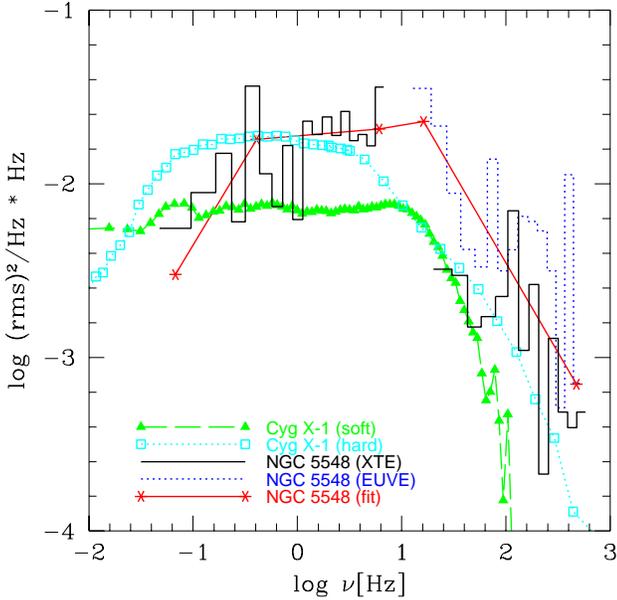}
\caption{Power density spectrum times frequency for NGC 5548 
from Chiang et al. (2000) (histogram: continuous line - XTE, 
dotted line - EUVE; stars connected with continuous line - fit) and
for galactic black holes: Cyg X-1 in hard state from Revnivtsev et
al. (2000) (open squares connected with dot line), Cyg X-1 in soft
state from Gilfanov et al. (solid triangles connected with long dash
line). Power density spectrum of NGC 5548 was shifted by 6.83 in
$log(\nu)$ which corresponds to rescaling it to a black hole mass of
$10 M_{\odot}$.
\label{fig2}}
\end{figure}

Time variability of the X-ray and optical emission from the nucleus of
NGC 5548 was studied by a number of authors (e.g. Papadakis \&
Lawrence 1993, Clavel et al. 1992, Czerny, Schwarzenberg-Czerny
\& Loska 1999, Chiang et al. 2000 and references therein). 
The power density spectrum is basically featureless, as in other AGN,
and the variability is caused by some stochastic process (Lawrence et
al. 1987, McHardy \& Czerny 1987, Czerny \& Lehto 1997).  Since
Comptonization is directly responsible for the formation of the hard
X-ray spectra this variability must be related to variable seed photon
flux and/or hot plasma properties but the exact physical mechanism is
still unknown (for a review of X-ray spectra and variability of AGN,
see e.g. Mushotzky, Done \& Pounds 1993, Nandra et al. 1997, Leighly
1999a,b; but see also Abrassart \& Czerny 2000 and the references
therein for an alternative view). The same mechanism is expected to
operate in galactic black holes (GBH; for a review, see e.g. van der
Klis 1995, Cui 1999, Poutanen 2000).

The basic similarity of AGN and GBH with respect to the spectra
and variability was discussed by a number of authors (e.g. Zdziarski 1999).
There are, however, also some systematic differences reflecting the
direct dependence on the mass of the central body: (i) the temperature
of the disc component is about two orders of magnitude higher in
galactic sources than in AGN (ii) GBH in their soft/high state are
generally less variable than AGN dominated by the disc component.

Recent study of NGC 5548 with RXTE satellite allowed to determine an 
exceptionally accurate PDS for this source in the X-ray 
band (Chiang et al. 2000).

We compare PDS of NGC 5548 with PDS of a number of galactic sources in
various luminosity states (see Fig.~\ref{fig1}). We plot these spectra
in the form of $\nu \times power$ for more convenient comparison. We
rescale the PDS of NGC 5548 to the mass of the black hole $10
M_{\odot}$, appropriate for Cyg X-1 (see Nowak et al. 1999 and the
references therein) through simple horizontal shift by a factor of
$6.8 \times 10^6$. We show separately the comparison of NGC 5548 with
the high quality data of Cyg X-1 in its high/soft (Gilfanov et
al. 2000) and low/hard (Revnivtsev, Gilfanov \& Churazov 2000) states
(see Fig.~\ref{fig2}).

The exact comparison of the two distributions is however difficult
since the quality of the PDS for NGC 5548 is much lower than that of
Cyg X-1. The fit to the overall shape of the PDS found by Chiang et
al. (2000) suggests that there is a systematic shift between the two
distributions (see continuous line in Fig.~\ref{fig2}) by about 0.4 in
the logarithmic scale.  The high frequency turn off is best reproduced
by Cyg X-1 in soft or in the transition state, but the normalization
is considerably higher, confirming much stronger variability than in
the soft state of Cyg X-1.  On the other hand the fit is strongly
influenced by the results from EUVE which represents different energy
band.  XTE results are roughly consistent with the distribution of Cyg
X-1 in the high frequency part.

The exact position of the high frequency part of the spectrum is the
basis of the mass measurement proposed by Hayashida et al. (1998).
The dependence on the luminosity is constrained to low frequencies
(Belloni \& Hasinger 1990) so the high frequency part (above 10 Hz for
galactic sources and correspondingly lower for AGN) is promising for
mass determination.

According to Hayashida et al., the frequency $\nu_{0.001}$ where the
power spectrum in $\nu \times Power$ representation is equal $10^{-3}$
scales with the mass, ${\rm log} (M_{BH}/M_{\odot}) = C - {\rm log}
\nu_{0.001}$. Hayashida et al. (1998) adopted the value $C$ equal 2.66
at the basis of their Cyg X-1 spectrum. The new hard state data
indicates the value of $\nu_{0.001}$ by more than a factor of 2
higher, ($C = 3.1$).

The power spectrum of NGC 5548, if well represented by the fit, is
slightly shifted towards higher frequencies and if the mass of
Peterson \& Wandel (1999) is adopted, $C = 3.5$.

The difference between the value of $C$ from Cyg X-1 and NGC 5548 is
larger than the formal error given on the mass measurement of NGC 5548
(0.1; Peterson \& Wandel 1999).

This difference may be partially caused by the poor quality of NGC
5548 data.  Simulations indicate that the determination of the high
frequency slope of the PDS may be biased by the power leak to high
frequencies connected with the window function if the PDS slope is
close to -2 (Green, McHardy \& Lehto 1993; see also Czerny et
al. 1999).

The shift may also reflect the uncertainty in our knowledge of the
masses: neither the mass of Cyg X-1 nor of NGC 5548 is known very
accurately, if systematic errors connected with reverberation method
are taken into account (Krolik 2000).  From this point of view, it
would be more convenient to use the galactic source GRO J1655-40
because the mass determination for this source is by far the most
accurate ($ M = 7.2 \pm 0.22 M_{\odot}$, Orosz \& Bailyn 1997; see the
discussion of Zi\' o\l kowski 2001). Unfortunately, the currently
available power spectrum for this source (Sunyaev \& Revnivtsev 2000)
is not as accurate as for Cyg X-1 so at the basis of this source the
value of $C$ should be roughly between 2.9 and 3.3, supporting the
result from Cyg X-1.

The systematic shift between NGC 5548 and Cyg X-1 comes predominantly
from the EUVE results biasing the fit proposed by Chiang et
al.(2000). The XTE results themselves (a histogram in Fig.~\ref{fig2})
are roughly consistent with Cyg X-1.

Therefore, assuming the  value $C=3.1$ for both galactic objects and AGN
seems reasonable. 


\section{Determination of the central black hole masses in AGN} 
\label{bhmasy}

\begin{figure}
\epsfxsize = 90 mm 
\epsfbox{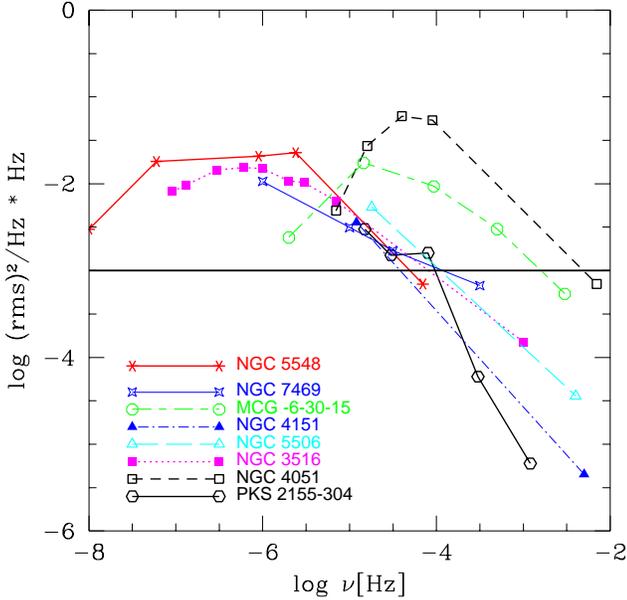}
\caption{Power density spectrum times frequency for NGC 5548 (stars 
connected with continuous line), PKS 2155-304 (open hexagons connected
with continuous line), NGC 3516 (solid squares connected with dot
line), NGC 4051 (open squares connected with short dash line), NGC
4151 (solid triangles connected with dot-short-dash line), NGC 5506
(open triangles connected with long dash line), NGC 7469 (solid
hexagons connected with short dash line), MCG -6-30-15 (open octagons
connected with short-dash-long-dash line). PDS of these objects are
used from papers cite in text (see Section 3).
\label{fig3}}
\end{figure}

We use the available PDS for a number of Seyfert galaxies in order to 
determine their masses: PKS 2155-304, NGC 4051, NGC 4151 and NGC 5506
is taken from Hayashida et al. 1998, MCG -6-30-15 is from Nowak \& 
Chiang (2000), and NGC 3516 from Edelson \& Nandra (1999).
PDS for NGC 7469 is from Nandra \& Papadakis (2001),
but with the normalization different by a factor of 2 from the original
normalization which assures that the integral of the PDS gives the variance.
All distributions, together with NGC 5548, are plotted in Fig.~\ref{fig3}.

The PDS of various AGN are essentially similar to each other and to the
PDS of NGC 5548. Horizontal shifts between them can be explained as the
differences in their black hole masses. However, there are also some 
differences in total normalization. Some of the effect may be connected
with the finite length of the data, not extending beyond the maximum of
the PDS at the longest timescales. However, the effect may be also 
connected with their luminosity. Similar dispersion is also seen among the
galactic sources. It clearly limits the accuracy of the method.

We determine the black hole masses from their PDS using the method of
scaling adopted by Hayashida et al. (1998). We adopt the PDS value of
$10^{-3}$(rms)$^2$ Hz$^{-1}$Hz as the characteristic value and we
compare the frequencies $\nu_{0.001}$ where these value is reached in
various sources (thick horizontal line in Fig.~\ref{fig3}).
Our formula is (see Sect.~\ref{sect:5548vsgbh})
\begin{equation}
{\rm log} (M_{BH}/M_{\odot}) = 3.1 - {\rm log} \nu_{0.001}.
\end{equation}

For those sources analyzed by Hayashida et al. (1998) and Hayashida
(2000) for which no better PDS are currently available we simply adopt
their results after scaling the masses by a factor 2.8 (shift by 0.44
in logarithm), which results from systematic difference between new
and previous results for Cyg X-1 (see Sect.~\ref{sect:5548vsgbh}). 
Those sources are: PHL 1092, PG 1244, IRAS 13224, 1H 0707 
from Hayashida (2000), and 3C 273, ESO 103-G35, PKS 2155-304, NGC 4051, 
NGC 4151 and NGC 5506 from Hayashida et al. (1998).

All masses of central black holes, are given in 
Table~\ref{tab1}.


\begin{table*}
\caption{Properties of AGN.
Black hole masses are determined from variability
(Section~\ref{bhmasy}), photoionization and reverberation method
(Wandel, Peterson \& Malkan 1999), accretion disk fits
(Section~\ref{diskmasy}), and the first value is used to calculate the
first luminosity to the Eddington luminosity ratio and the mass value
from the disk fits are used to calculate the second ratio. Last column
gives the mass of the bulge.  Units: $M$ is in $M_{\odot}$ and $L$ is
in $erg\; s^{-1}$.
\label{tab1}}
\begin{tabular}{llllllllll}
\hline\hline
\\
object & type & ${\rm log} M_{BH}^{var}$ & ${\rm log} M_{BH}^{ph}$ & ${\rm log} M_{BH}^{rev}$ & 
${\rm log} M_{BH}^{disk}$ & ${\rm log}L_{bol}$ & $ L/L_{Edd}^{var}$ & $ L/L_{Edd}^{disk}$ & 
${\rm log} M_{bulge}$\\
\hline
\\
NGC 5548        &S1.5&  7.39      &      &      & 7.31 & 44.36 & 0.074   & 0.089 &  11.47   \\
NGC 7469        &S1.2&  7.03      & 6.87 & 6.88 & 7.41 & 44.30 & 0.15    & 0.062 &  11.03   \\
MCG -6-30-15    &S1.2&  5.94      &      &      & 6.98 & 43.85 & 0.646   & 0.059 &  10.23   \\
NGC 4151        &S1.5&  7.57      & 7.35 & 7.08 & 7.03 & 43.70 & 0.010   & 0.037 &  11.05   \\
                &   &             &      &      & 7.33 & 43.93 & 0.018   & 0.032            \\
NGC 5506        &S1.9&  6.86      &      &      & 7.23 & 43.88 & 0.083   & 0.036 &  10.78   \\
ESO 103-G35     &S1.2&  7.14      &      &      & 7.44 & 43.39 & 0.014   & 0.007 &  10.62   \\
NGC 3516        &S1.5&  7.25      &      & 7.30 & 7.49 & 44.46 & 0.13    & 0.074 &  11.47   \\
NGC 4051        &NLS1&  5.39      & 5.37 & 6.15 & 6.50 & 42.40 & 0.082   & 0.006 &  10.91   \\
IRAS 13224-3809 &NLS1&  4.52      &      &      & 8.01 & 44.95 & 215     & 0.069 &  12.16   \\
1H 0707-495     &NLS1&  5.61      &      &      & 6.77 & 43.69 & 0.95    & 0.066 &  11.39   \\
                &    &            &      &      & 6.85 & 44.60 & 7.73     & 0.45            \\
PG 1244+026     &NLS1&  5.85      &      &      & 6.62 & 44.13 & 1.51    & 0.26 &   11.42   \\
PHL 1092        &NLS1&  6.09      &      &      & 8.26 & 45.70 & 32      & 0.22 &    ---    \\
3C 273          &QSO &  8.12      &      &      & 8.61 & 46.58 & 2.3    & 0.74 &   12.39   \\
PKS 2155-304    &QSO &  7.39      &      &      & 8.12 & 46.21 & 5.26    & 0.98 &    ---    \\
\hline
\end{tabular} \\
\end{table*}

\section{Bulge masses of host galaxies}

We determine the bulge masses of the considered AGN using the data
available from the literature and following the approach of Laor
(1998) and Wandel (1999).

No data are available on the host galaxy of PHL 1092 and PKS 2155-304,
so we were unable to determine bulge masses of host galaxies in the
case of these two objects.

For NGC 5548, NGC 4051, NGC 3516, NGC 4151, NGC 5506 we take the blue
bulge magnitude $M_B$ from Whittle (1992), we calculate the visual
magnitude as $M_V = M_B - 0.95$ (Worthey 1994), bulge luminosity as
${\rm log}(L_{bulge}/L_{\odot}) = 0.4(-M_V + 4.83)$ and finally ${\rm
log} (M_{bulge}/M_{\odot})= 1.18{\rm log}(L_{bulge}/L_{\odot}) -
1.11$, after Magorrian et al. (1998).  Our values of masses for 5548,
NGC 4051, and NGC 4151 differ by a factor of $\sim 2$ from those of
Wandel (1999) due to the $M_V = M_B - 0.95$ correction applied by us.

For 3C 272 we take the visual magnitude of the host galaxy
$M_V^{total}=-22.58$ from Laor (1998), and the type of the galaxy E4
after Bahcall et al. (1997), we then calculate the bulge luminosity
from the formula $M_V^{total} - M_V= 0.324 \tau - 0.054 \tau^2 +
0.00047 \tau^3$, where $\tau = T+5$ and $T$ is the Hubble type of the
galaxy and we further proceed as described above.

In the case of remaining AGN, apart from IRAS 13224-3809, we took
their apparent magnitudes and redshifts from NED, we corrected for the
contamination of the total flux by the nuclear emission using the
measured equivalent width of the $H_{\beta}$ line, $EWH_{\beta}$, as
$M_B^{total}=M_B^{measured} - 2.5 {\rm log}(1 - 1.10*(1 +
z)EWH_{\beta}/100)$, after Whittle (1992), assuming the spectral slope
equal -1 in all objects.  $EWH_{\beta}$ values were taken from Morris
\& Ward (1988) for MCG -6-30-15, and ESO 103-G35, and from Miller et
al. (1992) for PG 1244+226, and from Giannuzzo \& Stirpe (1996) for 1H
0707-495.  We also included the correction for intrinsic absorption,
0.14, the same in all sources, after Whittle (1992). The host galaxy
luminosity of IRAS 13224-3809 was estimated on the basis of V band
magnitude of the active nucleus 15.2 from Young et al. (1999) and
large aperture value of $EWH_{\beta}$ equal to 23 \AA~ given by Boller
et al. (1993).

We see that S1 galaxies and 3C 273 follow the bulge mass--black hole
mass relation determined by Laor (1998) and Wandel (1999) but they are
clearly shifted from the Magorrian et al. (1998) relation for low mass
objects. Formally, in our data the relation holds
\begin{equation}
M_{bulge} \propto M_{BH}^{var \, 0.98},
\end{equation}
if the black hole mass is determined from the variability (Sect.~\ref{bhmasy}).

The relation recently found by McLure \& Dunlop is close to the
original relation of Magorrian et al. (1998) and it is clearly
inconsistent with our results.

NLS1 galaxies are strongly shifted from the S1 distribution towards
low black hole mass range for approximately the same range of bulge
masses of host galaxies.

\section{Determination of the $ L/L_{Edd}$ ratio of AGN}
\label{sect:edd}

We determine the bolometric luminosity of AGN as a sum of two
independent contributions from X-ray and optical/UV band on the basis
of the integrated 2-10 keV flux and $\nu F_{\nu}$ luminosity at 2500
\AA~. We determine the bolometric luminosity at the basis of the
following formula: $ L_{bol} = 4.6 \times \nu L_{\nu}(2500 {\rm \AA})
+ 4.1 \times L(2-10 {\rm keV})$. Adopted bolometric corrections were
determined at the basis of the broad band spectrum of a Seyfert galaxy
NGC 5548 (Magdziarz et al.  1998), although we did not include the
entire unobserved XUV range, and of a Narrow Line Seyfert 1 galaxy
PG 1211+143 (Janiuk, Czerny \& Madejski 2001). This approach seems to
be more appropriate than using only 2 -10 keV flux with a bolometric
correction 27.2 for all sources (Padovani \& Rafanelli 1988, Hayashida
et al. 1998) or using only the optical flux ($L_{bol}=9 \lambda
L_{\lambda}$ (5100 \AA), Kaspi et al. 2000) since Seyfert galaxies
clearly show a large range of the Big Blue Bump to the hard X-ray
power law luminosity ratios (Walter \& Fink 1993), particularly if
NLS1 galaxies are included in the sample.

The data for NGC 5548 was taken from Magdziarz et al. (1998), X-ray
2-10 keV fluxes for MCG -6-15-30, NGC 4051, NGC 4151, ESO 103-G35, NGC
5506, PKS 2155-304 and Cyg X-3 are from Hayashida et al. (1998), for
NGC 3516 from Nandra et al. (1997), and for IRAS 13224-3809, 1H
0707-495, PG 1244+226 and PHL 1092 are from Leighly (1999a).  UV
fluxes for MCG -6-15-30, and 3C 273 are from Walter \& Fink (1993), NGC
4051 is from Mathur et al. (2000), NGC 7469 from Nandra et al. (1998).
For PG 1244+026 the optical flux was taken from Miller et al. (1992),
and for 1H 0707-495 from Giannuzzo \& Stirpe (1996).  UV fluxes for
other sources were obtained from their V or B magnitudes given in NED.
For 1H 0707 we also used a second set of data coming from much older
observations when the source seemed to be fainter: optical flux and
2-10 keV flux were taken from Remillard et al. (1986). Second set of
data for NGC 4151 comes from the later observations in 1993: B
magnitude is taken from Lyuty \& Doroshenko (1999) and X-ray flux
from Weaver et al. (1994).

\section{Discussion} 

\subsection{Comparison with black hole mass determination 
from Broad Line Region}

The applicability of the method based on the Power Density Spectrum
can be tested using those objects which black hole mass was determined
independently.

Some of the sources in our study were extensively monitored (NGC 4151,
NGC 4051, and NGC 7469) which allowed to use reverberation or
photoionization method, or both, to determine the mass of the central
black hole. The values obtained by Wandel, Peterson \& Malkan (1999)
are given in Table~\ref{tab1}. NGC 3516 was also monitored and the
black hole mass determined by reverberation method was given by
Wanders \& Horne (1994).

We see that for three objects there is a good agreement between the
black hole mass determined from X-ray variability with the values from
both the ionization and reverberation method. Methods based on the
Broad Line Region properties are themselves accurate within a factor
of three, according to the discussion recently made by Krolik (2000),
and the mass from PDS is contained in the expected range.  In case of
NGC 4151 the black hole mass determined from variability is larger by
a factor of 1.7 than the value from photoionization and by a factor of
3 than the value from reverberation. In that case the power density
spectrum taken from Hayashida et al. (1998) may not be determined
accurately. This object, having larger mass, requires longer
monitoring than NGC 4051 and the Ginga results may not be
adequate. The source was earlier observed by EXOSAT and the power
spectrum was determined by Papadakis \& McHardy (1995) but without the
normalization. If we normalize their spectrum according to the
provided mean intensity and the excess variance and use their PDS we
obtain even larger black hole mass (log $M_{BH} = 8.43$). However,
true high frequency part of the PDS may be actually flatter than the
slope -2 given in the paper due to the power leakage effect discussed
by the authors. If we arbitrarily assume that the turning point is
correct but the slope is -1.5 we obtain log $M_{BH} = 7.5$. This last
value is closer to the values determined from the BLR properties, but
this discussion mostly points out a need of high quality PDS of AGN.

\subsection{Comparison with  black hole mass determination from accretion disk fitting}
\label{diskmasy}

\begin{figure}
\epsfxsize = 90 mm 
\epsfbox{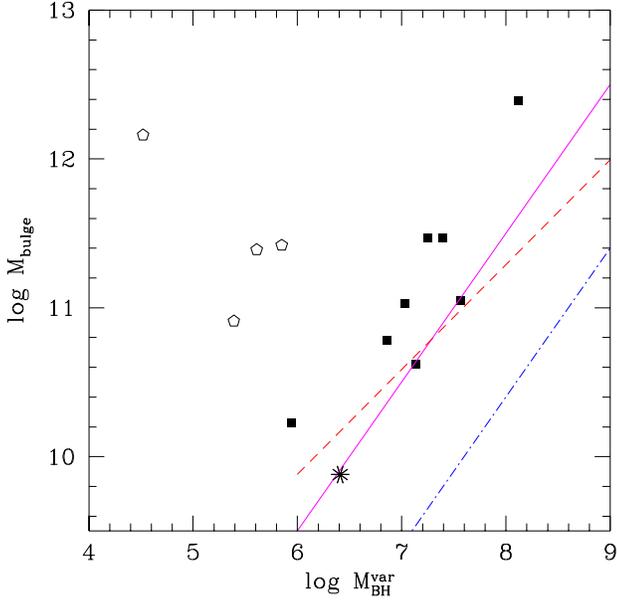}
\caption{The relation between the mass of the central black hole 
$M_{BH}^{var}$ determined from variability and the mass of the host
galaxy bulge: filled squares show S1 galaxies and 3C 273, and open
pentagons mark NLS1. Continuous line shows the relation of Wandel
(1999) for Seyfert galaxies, dashed line is for quasars and normal
galaxies (Laor 1998), and dot-short-dash line is the original relation
of Magorrian et al. (1998). Star marks the position of the Milky Way.
\label{figmass}}
\end{figure}

For most of the sources from our sample such determinations are not
available (and for other objects from Wandel et al. (1999) sample the
power spectra were not determined). Therefore, in order to test better
the method we determine the black hole masses for all sources based on
the following assumptions: (1) outer part of the flow responsible for
the UV emission is well modeled by the standard accretion disk (2) the
X-ray emission is produced closer in, and the total efficiency of
conversion of the accreting mass into observed radiation is 1/16, as
appropriate for non-advecting flow onto a Schwarzschild black hole.

The first assumption results in adopting the relation between the
black hole mass, accretion rate, intrinsic luminosity $L_{\nu}$ at 
a frequency $\nu$ after Tripp, Bechtold \& Green (1994)
\begin{equation}
\label{eq:trip}
log M + log \dot M = 1.5 (log L_{\nu} - 19.222) - 0.5 log \nu,
\end{equation}
where $M$ is in solar masses, $\dot M$ is solar masses per year, $\nu$
in Hz, and $L_{\nu}$ in erg s$^{-1}$ Hz$^{-1}$.

This relation, combined with the second assumption and calculated at
2500 \AA, gives
\begin{equation}
log M = 1.5 log(\nu L_{\nu}^{2500}) - log L_{bol} - 13.44.
\end{equation}
Since it relies on disk assumption we call it $M_{BH}^{disk}$ in
Table~\ref{tab1}.

Such an approach does not take into account the departure of the
locally emitted spectrum from the black body, and advection or outflow
may lead to smaller energetic efficiency $\eta$ than assumed.  The
resulting black hole mass depends on those factors as
\begin{equation}
M \propto f^2 \eta
\end{equation}
(Janiuk et al. 2001), where $f$ is the color temperature to the effective 
temperature ratio so the mass may be either underestimated or overestimated
by a factor of a few.

\begin{figure}
\epsfxsize = 90 mm 
\epsfbox{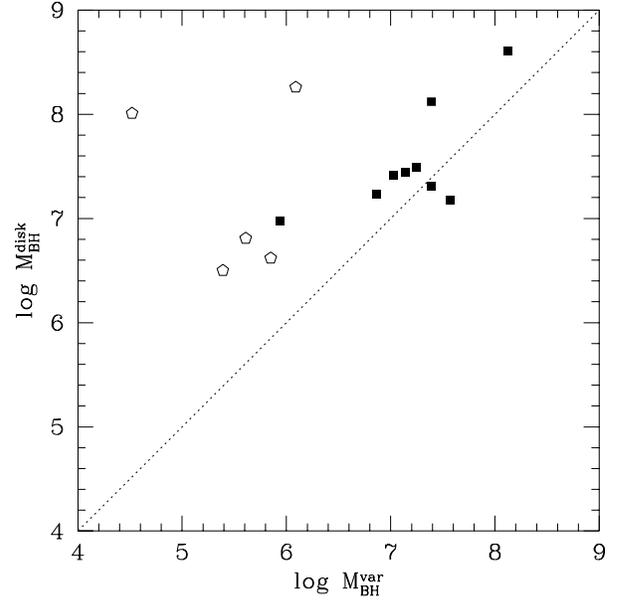}
\caption{The relation between the mass of the central black hole $M_{BH}^{var}$
determined from variability and the mass determined on the basis of
accretion disk theory $M_{BH}^{disk}$: filled squares show S1 galaxies and
QSOs, and open pentagons mark NLS1. Dotted line marks the expected
relation $M_{BH}^{var} = M_{BH}^{disk}$.
\label{fig:mama}}
\end{figure}

The relation between the masses determined in two ways are shown in 
Fig.~\ref{fig:mama}.

The black hole masses obtained in this way for S1 galaxies and QSOs
are quite similar to values obtained from variability. Mean value of
the logarithm of the black hole mass for a S1 galaxy is equal to 7.03
from variability and 7.27 from disk fitting, and for the two quasars these 
two numbers are equal 7.75 and 8.36, correspondingly. It might indicate
a slight systematic non-linearity required in the formula connecting the
mass with the frequency $\nu_{0.001}$ (see Section~\ref{sect:mass5548}) 
but the uncertainties
are currently too large to address this issue. 

Recent paper of Collin \& Hure (2001) claimed a considerable problem with
accretion disk luminosity but we did not found any evidence for such an effect
(see Appendix A).

However, we note a large discrepancy between the methods in the case
of NLS1, with mean masses equal 5.49 from variability and 7.16 from
disk fitting. The power spectra for four out of five objects comes
from Hayashida (2000) and they are of low quality, as indicated by
large errors given by the author, but better spectra are not
available. Higher quality spectrum available for NGC 4051 gives the
value of the mass located between the two values provided by the BLR
studies which may indicate that the scaling with NGC 5548 which is S1
may be nevertheless appropriate for NLS1 objects. In this case a
larger error may be actually connected with the way how we determined
the bolometric luminosity which influenced both the luminosity to the
Eddington luminosity ratio as the determination of the
$M^{disk}_{BH}$. Our formula (given in Sect.~\ref{sect:edd}) is
appropriate for sources either strong in X-rays (as S1 galaxies) or
sources with a Big Blue Bump having a maximum in $\nu F_{\nu}$ in UV
band, like PG 1211+143 or quasar composite spectrum of Laor et
al. (1997). However, for some NLS1 the Big Blue Bump may be located in
very far UV or even soft X-rays. One example of such a source (RE
J1034+396) was found by Puchnarewicz et al. (1995). In this extreme
case the bolometric luminosity estimated from our formula is $6.4
\times 10^{43}$ erg s$^{-1}$, while direct integration of the broad
band spectra would give about $ 2 \times 10^{45}$ erg s$^{-1}$, a
factor of 30 higher, and the first value would give a black hole mass
by a factor of 30 too high.  Large error is therefore possible,
particularly for sources which are weakly variable in the optical band
like NGC 4051, which may also have a Big Blue Bump shifted
considerably towards higher energies, although not as high as in RE
J1034+396. Therefore, $M^{disk}_{BH}$ may be sometimes too high while
on the other hand the values of $M^{var}_{BH}$ for IRAS 13224-3809, 
1H 0707-495 and
PHL 1092 are clearly too low, leading to extreme super-Eddington
luminosities. 

Recent results for the X-ray variability of NLS1 galaxy Akn 564 does
not solve the issue since only the rising and the flat part of the
power spectrum on $\nu \times Power$ diagram were detected (Pounds et
al. 2001) thus giving only a lower limit on the frequency
$\nu_{0.001}$ ($\sim 10^{-4}$ Hz and subsequently an upper limit on
the black hole mass ($10^7 M_{\odot}$) since the monitoring procedure
did not allow to study higher frequencies.

Summarizing, the PDS method seems to lead to good black hole mass
estimation for normal Seyfert galaxies and QSOs the high quality PDS
is available. The conclusion is less firm for NLS1 but this is
probably caused by low quality of the available PDS spectra for those
objects.

\subsection{Luminosity to the Eddington luminosity ratio}

The luminosity to the Eddington luminosity ratio for S1 galaxies is
$\sim 0.05$ up to $0.07$, practically independently from the method 
used to determine the masses of their central black holes.

Only one source, MCG -6-30-15, has this ratio higher than 0.15.  This
object is rather exceptional with respect to the properties of its
X-ray spectra: it has relatively broad iron $K_{\alpha}$ line, with a
strong red wing (Tanaka et al. 1995, Nandra et al. 1997). MCG -6-30-15
has also very soft spectrum, with a hard X-ray photon index in ASCA
data 1.95 and 2.04 in two data sets (Nandra et al. 1997).
The $H_{\alpha}$ line of MCG -6-15-30 is rather narrow, about 2000 km
s$^{-1}$, and the $H_{\beta}$ line also seems quite narrow (Morris \&
Ward 1988), although the exact value may strongly depend on the
decomposition of the line into narrow and broad component.  Turner et
al. (1999) quote the value of 1700 km s$^{-1}$ after Pineda et
al. (1980).  Classification of this source should be perhaps
reconsidered.

NLS1 galaxies show systematically higher luminosity to the Eddington
luminosity ratios (although the strength of this effect depends on the
black hole mass measurement method). This is consistent with suggestion
of Pounds, Done \& Osborne (1995), confirmed by a number of later
studies (e.g. Czerny, Witt \& \. Zycki 1996, Wandel 1997, Brandt \&
Boller 1998, Kuraszkiewicz et al. 2000).

Black hole mass estimate based on the variability method leads for
three sources to unrealistic, highly super-Eddington values of the
accretion rate. Higher quality power spectra for these sources are
clearly needed. Spectral fitting approach lead to much smaller ratios,
with mean value 0.18, more than three times higher than for S1
galaxies. NLS1 galaxies have on average higher luminosity to the
Eddington luminosity ratio although some NLS1 galaxies have those
ratios (usually only slightly) smaller than some of S1 galaxies.

Only NGC 4051 among NLS1 has an exceptionally low luminosity to the
Eddington luminosity ratio.  The mass determined from the variability
seems to be correct for this source, as it is intermediate between the
other values quoted in the literature: $H_{\beta}$ study gave $(1.1 -
1.4) \times 10^6 M_{\odot}$, (Peterson et al. 2000), Kaspi et
al. (2000) obtained $(1.3 - 1.4) \times 10^6 M_{\odot}$, and Wandel et
al. (1999) values were $2.3 \times 10^5M_{\odot}$ and $1.4 \times 10^6
M_{\odot}$, as given in Table~\ref{tab1}. It may be that the source
bolometric luminosity is by an order of magnitude underestimated if
the broad band spectrum is dominated by the unobserved XUV range. On
the other hand Peterson et al. (2000) argue that this source may have
relatively narrow $H_{\beta}$ line due to high inclination ($\sim 50
^o$, Christopoulou et al. 1997) instead of high $L/L_{Edd}$. Better
determination of the PDS for this source may resolve this issue.

Two QSO included in our sample have rather high $ L/L_{Edd}$ ratios.
The black hole mass of 3C 273 based on variability and on our simple
spectral fit is smaller than the value determined from detailed
spectral fitting (e.g. Leach, McHardy \& Papadakis 1995).  The problem
may be connected with too short monitoring, i.e.  the lack of coverage
of frequencies close to flattening point and subsequently, a problem
with normalization of the PDS. On the other hand, in the Leach et
al. (1995) fits the energy required to heat the Comptonizing corona is
not included in the computation of the accretion rate, possibly giving
too low accretion rate and too high mass.  The analysis of longer
timescale variability in these sources is clearly needed.

\subsection{Spectral states of AGN}

There are considerable similarities between the accretion process onto
stellar black holes and supermassive black holes: similar range of
X-ray spectral slopes (Zdziarski 1999), similar correlation between
the normalization of the reflected component and X-ray spectral slope
(Zdziarski, Lubi\' nski \& Smith 1999) and basically similar
featureless PDS (Hayashida et al. 1998). The shape of X-ray PDS for
both NLS1 galaxies and normal S1 galaxies look quite similar to the
GBH in their hard state.

Detailed comparison of AGN and GBH shows, however, a systematic
difference: (i) GBHs in their soft state show very little
variability, with rms at a level of a few per cent (van der Klis 1995)
while even disc-dominated AGN like quasars generally show higher
variability both in optical (Giveon et al. 1999) and X-ray band
(e.g. Yaqoob et al. (1994) for PG 1211). 

\begin{figure}
\epsfxsize = 90 mm 
\epsfbox{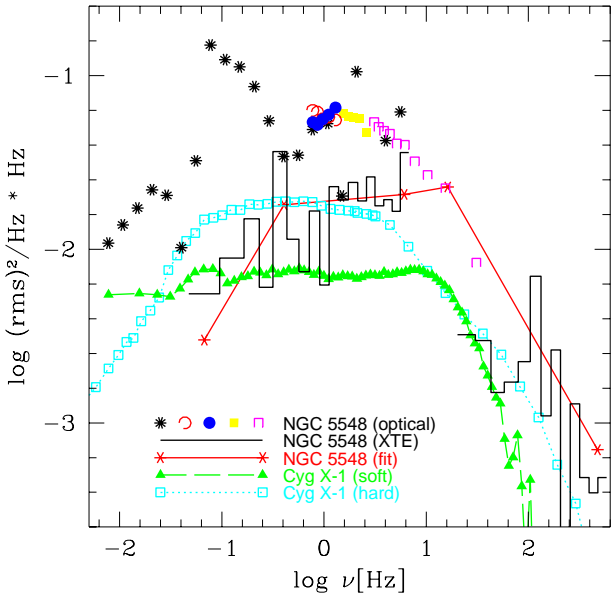}
\caption{The comparison of the optical power density spectrum times 
frequency for NGC 5548 (Czerny et al. 1999; disconnected points) with
the X-ray power density spectrum of NGC 5548 from Chiang et al. (2000)
(histogram: continuous line - XTE; stars connected with continuous line
- fit) and for galactic black holes: Cyg X-1 in hard state from
Revnivtsev et al. (2000) (open squares connected with dot line), Cyg
X-1 in soft state from Gilfanov et al. (solid triangles connected with
long dash line). Power density spectra of NGC 5548 were shifted by
6.83 in $log(\nu)$ which corresponds to rescaling them to a black hole
mass of $10 M_{\odot}$.
\label{opt}}
\end{figure}

This apparent inconsistency can be qualitatively understood in the
following way.  The similarity of the X-ray PDS for all kinds of AGN
reflects the fact that this PDS is usually calculated using the 2 - 10
keV energy band which is always dominated by the hard X-ray power
law. X-ray emission of the galactic sources in their soft state is,
however, dominated by the soft component which roughly corresponds to
the optical/UV component in AGN.

The same variability mechanism responsible for the X-ray PDS of AGN
and GBH in their hard states does not mean that the two must be
identical. Actually, we would rather expect slight systematic shift
between AGN and GBH (apart from the one resulting from the mass ratio)
since the observed PDS of galactic sources show some dependence of PDS
on the studied spectral range (e.g. Cui et al. 1997, Nowak et
al. 1999, Gilfanov, Churazov \& Revnivtsev 1999).  The spectrum of NGC
5548 is also dominated by hard X-ray power law, without a strong
contribution from soft X-ray excess. However, soft seed photons for
Comptonization have much lower energy in case of AGN and have to
undergo more scatterings before reaching the same energy as photons in
Cyg X-1. Therefore, even if the hot medium temperature and geometry is
identical in an AGN and a GBH the Comptonization process is not
identical and an AGN PDS may resemble more the PDS of a GBH at
somewhat higher energy band. Numerical simulations should be used to
study such a scaling properties with the black hole mass, i.e. with
the soft photon temperature.

The essential differences between AGN and GBH are reflected in the
behavior of their soft spectral components. Cyg X-1 in disk-dominated
soft state shows very moderate variability. The behavior of disk
component in NGC 5548 is well represented by the {\it optical} PDS
shown in Figure~\ref{opt}.  The errors in determination of the optical
PDS are large but we see that the overall shape is more similar to
that of Cyg X-1 in its hard state than the X-ray PDS. It may mean that
the basic variability mechanism is similar in both objects and that
the spectral state of NGC 5548 really corresponds to low/hard state,
as suggested by its $L/L_{Edd}$ ratio, although the normalization of
the optical PDS of NGC 5548 is much higher than that of Cyg X-1.

Such an effect is not surprising. According to a standard
Shakura-Sunyaev disc model (Shakura \& Sunyaev 1973) we deal 
with the different ratio of the radiation to the gas pressure in
accretion discs in AGN and GBH at the same $L/L_{Edd}$ ratio.  Time
evolution and hot plasma formation rate can clearly depend on this
ratio.  R\' o\. za\' nska \& Czerny (2000) show the expected
difference between AGN and GBH accretion flow stability in the case of
accreting corona model.  In this case an evaporation branch exists
which limits optically thick solutions to high accretion rates only,
with a transition to an advection-dominated accretion flow (i.e. ADAF;
Ichimaru 1977, Narayan \& Yi 1994, for a review see Narayan, Mahadevan
\& Quataert 1998) for low accretion rates. The accretion rate at which
the evaporation branch forms does not depend on the mass of the black
hole but the extension of the gas pressure dominated branch of a
standard disc does depend on the mass considerably.  As a result, AGN
never achieve a stable high state since before the accretion rate
drops enough to allow the gas pressure to dominate the disc
evaporates.

\subsection{Black hole to galactic bulge mass ratio}

\begin{figure}
\epsfxsize = 90 mm 
\epsfbox{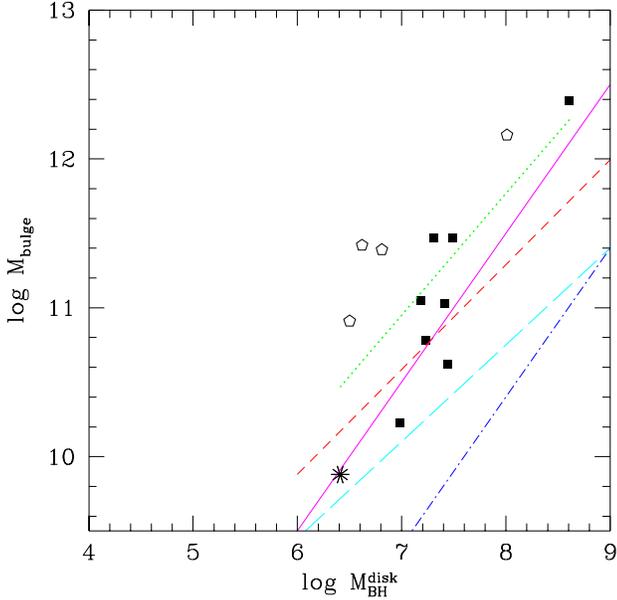}
\caption{The relation between the mass of the central black hole determined
from spectral estimates ($M_{BH}^{disk}$ from Table~\ref{tab1}) and
the mass of the host galaxy bulge: filled squares show S1 galaxies and
3C 273, and open pentagons mark NLS1. Continuous line shows the
relation of Wandel (1999) for Seyfert galaxies, dashed line is for
quasars and normal galaxies (Laor 1998), dot-short-dash line is the
original relation of Magorrian et al. (1998), and long dash line is of
Laor (2001).  Dotted line shows our relation based of equation
(\ref{eqdiskmass}).  Star marks the position of the Milky Way.
\label{figdiskmass}}
\end{figure}

The mass of the central black hole residing in local massive galaxies
is well correlated with the mass of the bulge (Magorrian et al. 1998).
It suggests a close relation between the host galaxy and the formation,
or evolution, of its central black hole. Laor (1998) showed that nearby
quasars studied by Hubble Space Telescope follow the same relation.

However, Wandel (1999) studied this relation for Seyfert galaxies
and found that they form a similar relation as normal galaxies and
quasars but systematically displaced towards small black hole masses.
Similar study performed by Mathur et al. (2000) for Narrow Line Seyfert
1 galaxies showed even further displacement towards small black hole
masses.

Our mass determination used in Fig.~\ref{figmass} was based on PDS
analysis so it was different from method used by Mathur et al. (2000)
and Wandel (1999), and mostly performed for other objects.  Our
results, however, confirm the trends found by those authors - Seyfert
galaxies follow basically the relation determined by Wandel but NLS1
galaxies depart from it systematically. The strength of the effect,
however, significantly depends on the adopted method of black hole
mass measurement for NLS1 galaxies: NLS1 are shifted by $\sim 1.5$
orders of magnitude in Fig.~\ref{figmass} (black hole mass from
variability) and it is marginally shifted by $\sim 0.5$ magnitude in
Fig.~\ref{figdiskmass} (black hole mass from spectral fits). We
therefore may even neglect the difference between NLS1 galaxies and S1
galaxies and look for a universal relation between the bulge mass and
black hole mass for our objects. Spectral black hole mass
determination gives
\begin{equation}
M_{bulge} \propto M_{BH}^{disk \, 0.82},   
\label{eqdiskmass}
\end{equation}
closer to the non-linear relation found recently by Laor
(2001). However, without NLS1 galaxies this relation remains linear
for our objects even in the case of the black hole mass determination
from spectral fits.

The quasar 3C 273 in our analysis follows the relation for Seyfert galaxies.

Recent study of normal galaxies seem to indicate that even for these
objects the black hole mass and bulge mass relation is not universal,
showing trends with the mass of the galaxy (Merritt \& Ferrarese
2000).

\subsection{Ionization instability in AGN}

Ionization instability, first found to be responsible for the
outbursts of cataclysmic variables (e.g. Meyer \& Meyer-Hoffmeister
1984, Smak 1984) operates also in binary systems containing accreting
neutron stars or black holes.  It develops in outer, partially ionized
part of the disc. Causing periodically temporary enhancement or
suppression of the accretion it leads to transient behavior of many
sources in timescales of years (e.g. King 1995).  It was suggested
that similar instability may operate in AGN (e.g. Clarke \& Shields
1989, Siemiginowska, Czerny \& Kostyunin 1996, Hatziminaoglou,
Siemiginowska \& Elvis 2000) although expected timescales are of
thousands to millions of years. Such instability may modify the
accretion rate in the central parts of the flow thus being responsible
for temporary change of the object status from highly active (NLS1) to
regular (Seyfert 1) and perhaps quiescence (LINERS) in similar
timescales. On the other hand it is not clear that this instability
operates in AGN since the self-gravity effects are extremely important
in the outer partially ionized parts of discs around massive black
holes, leading to significant modification of the disc structure in
comparison with discs in low mass objects (e.g. Hur\' e 1998). The
fact that NLS1 galaxies and normal Seyfert 1 galaxies have different
black hole mass to bulge mass ratio and therefore different
evolutionary status argues against the temporary transition between
NLS1, S1 and LINER stages and in favor of the self-gravity preventing
the ionization instability to operate in AGN. Further studies,
however, are needed to confirm this conclusion.

\section{Conclusions}

\begin{itemize}
\item high frequency part of the PDS is luminosity-independent and 
can be used to determine the mass of the central object
\item Cyg X-1 may be used  as a reference object for AGN although 
systematic differences between AGN and GBH in the appropriate
normalization of the PDS up to an order of 0.4 in logarithm are not excluded
\item NLS1 galaxies may possibly require another scaling than S1 galaxies
\item the $L/L_{Edd}$ ratio based on mass determination
from PDS has large values for NLS1 galaxies and lower values for most
BLS1 galaxies. The effect is still present, although
significantly reduced, if complementary black hole mass measurement
based on simple spectral fitting is adopted,
\item NLS1 galaxies seems to have systematically lower black hole masses
in comparison to bulge masses of the host galaxies which supports the
conclusion of Mathur (2000) about their different evolutionary status.
MCG -6-30-15 occupies its appropriate part of the diagram 
despite having relatively large $L/L_{Edd}$, 
\item ionization instability operating in cataclysmic variables and 
X-ray novae probably cannot be responsible for variations of accretion
rate corresponding to transitions between NLS1 and BLS1 state
over the periods of thousands - millions of years. This instability
may be suppressed in AGN by
self-gravity effect.         
\end{itemize}

\bigskip

\section*{Acknowledgments}
We thank Marat Gilfanov for providing us with the PDS files for Cyg
X-1 as published in Gilfanov et al. (2000) and Revnivtsev et al. (2000), 
Jim Chiang for PDS files for NGC 5548 from Chiang et al. (2000), 
Paul Nandra for his permission to use the power spectrum
of NGC 7469 and helpful comments, and Piotr \. Zycki for many helpful
discussions.  This research has made use of the NASA/IPAC
Extragalactic Database (NED) which is operated by the Jet Propulsion
Laboratory, California Institute of Technology, and it was supported
in part by grant 2P03D01816 of the Polish State Committee for
Scientific Research.

\bigskip

\section*{Appendix}

In their recent paper Collin \& Hure (2001) analyzed the objects from
the sample of Kaspi et al. (2000) and they found serious deficiency of
the putative accretion disk luminosity in the optical band. Therefore,
we checked our approach outlined in Section~\ref{diskmasy} against
this sample.

If we assumed the same prescription for the bolometric luminosity as
Kapis et al. (2000), i.e.
\begin{equation}
L_{bol} = 9 \nu L_{\nu} (5100 \mathrm{\AA}),
\end{equation}
determine the accretion rate at the basis of the accretion efficiency $\eta$
\begin{equation}
L_{bol} = \eta \dot M c^2,
\end{equation} 
and adopt the value of the black hole mass from either of the two fits
(mean or rms, Equation 9 and 10 of Kaspi et al. 2000), we can
calculate the expected monochromatic accretion disk luminosity at 5100
\AA, in order to compare it with the observed value.

Expressing $M$ in solar masses and $\dot M$ in solar masses per year we obtain 
from Tripp et al. (1994) (see Equation~\ref{eq:trip}) in case of mean mass fit
\begin{equation}
{\rm log} \nu L_{\nu}^{disk}(5100) = 1.03 {\rm log} \nu L_{\nu}^{obs}(5100)
- 1.22  - {2\over 3} {\rm log} (16 \eta), 
\end{equation}
and for rms fit
\begin{equation}
{\rm log} \nu L_{\nu}^{disk} (5100) = 0.94 {\rm log} \nu L_{\nu}^{obs}(5100)
+ 2.56  - {2\over 3} {\rm log} (16 \eta),
\end{equation}
and for sources at the typical luminosities $ \nu L_{\nu}^{5100} \sim 10^{44}$
erg s$^{-1}$ there is no discrepancy between the disk luminosity and the
actual luminosity, contrary to Collin \& Hure (2001), if $\eta = 1/16$ and
$cos i = 0.5$ is assumed, as we did in our consideration. Collin \& Hure (2001)
adopted $\eta = 0.1$ and $cos i = 1$ which partially although not fully 
explains their conclusion about too low disk luminosity. Applying the
top view may actually be correct; however in that case such a factor should 
be probably included in the formula for the isotropic bolometric luminosity 
as calculated from the observed flux which will weaken the dependence to the
power 1/3.

\ \\
This paper has been processed by the authors using the Blackwell
Scientific Publications \LaTeX\  style file.

\end{document}